\documentclass[journal=jctcce,manuscript=article]{achemso}

\usepackage{amsmath}
\usepackage{amssymb}
\usepackage{xcolor}
\usepackage{hyperref}

\graphicspath{{./}}

\author{Benjamin Pampel}
\author{Omar Valsson}
\email{valsson@mpip-mainz.mpg.de}
\affiliation{Max Planck Institute for Polymer Research, Ackermannweg 10, D-55128 Mainz, Germany}
\alsoaffiliation{Present Address: Department of Chemistry, University of North Texas, 1508 W Mulberry St, Denton, TX 76201, USA. E-mail: omar.valsson@unt.edu}

\title{Improving the Efficiency of Variationally Enhanced Sampling with Wavelet-Based Bias Potentials}

\begin{document}

\begin{abstract}
  Collective variable-based enhanced sampling methods are routinely used on systems with metastable states, where high free energy barriers impede proper sampling of the free energy landscapes when using conventional molecular dynamics simulations. One such method is variationally enhanced sampling (VES), which is based on a variational principle where a bias potential in the space of some chosen slow degrees of freedom, or collective variables, is constructed by minimizing a convex functional. In practice, the bias potential is taken as a linear expansion in some basis function set. So far, primarily basis functions delocalized in the collective variable space, like plane waves, Chebyshev, or Legendre polynomials, have been used. However, there has not been an extensive study of how the convergence behavior is affected by the choice of the basis functions. In particular, it remains an open question if localized basis functions might perform better. In this work, we implement, tune, and validate Daubechies wavelets as basis functions for VES\@. The wavelets construct orthogonal and localized bases that exhibit an attractive multiresolution property. We evaluate the performance of wavelet and other basis functions on various systems, going from model potentials to the calcium carbonate association process in water. We observe that wavelets exhibit excellent performance and much more robust convergence behavior than all other basis functions, as well as better performance than metadynamics. In particular, using wavelet bases yields far smaller fluctuations of the bias potential within individual runs and smaller differences between independent runs. Based on our overall results, we can recommend wavelets as basis functions for VES.

\end{abstract}


\section{Introduction}

A major problem impeding conventional molecular dynamics~(MD) simulations is the so-called time scale or rare event problem. Often, the molecular process of interest occurs on a much longer time scale than one can simulate in practice; in other words, it is a rare event. Thus, the system stays in a metastable state during the simulation, and one does not observe transitions to other metastable states. Despite impressive developments in specialized hardware~\cite{Dror2012_BiomolecularSimulation,Anton3} and MD codes~\cite{NAMD_2020,GromacsGPU_2020} that make very efficient usage of modern graphics processing units, it is unlikely that accessible time scales will increase significantly in the near future. The speedup of individual processing units has come to an end and high-performance computing relies on the usage of massive parallelization~\cite{khan_science_2018}, and time is not easily parallelizable. Thus, there has been considerable interest in developing advanced methods that enhance phase space sampling and overcome this time scale problem~\cite{DicksonDinner_NonequilibriumEnhanced_2010,chong_pathsampling_2017,zuckerman_weighted_2017,husic_markov_2018,allison_computational_2020,kamenik_enhanced_2021,henin_enhanced_2022}.

A popular class of such advanced sampling methods is so-called collective variable (CV) based enhanced sampling methods. In these methods, we identify a few relevant coarse-grained order parameters, that is, CVs, that correspond to essential slow degrees of freedom. Typically, the selection of CV is made manually by using physical and chemical intuition~\cite{Fiorin_UsingCVs_2013,Giberti_MetaD_Crystals_2015,pietrucci_strategies_2017}, and sometimes requires a bit of trial and error, while methods based on machine learning are also showing great promise in automating this task~\cite{wang2020machine,noe2020machine,Gkeka2020mlffcgv,sidky2020machine}. The slow molecular process of interest is then associated with free energy barriers separating metastables on the free energy surface (FES) as a function of the chosen CVs. We then enhance the sampling of the FES by introducing an external bias potential that is adaptively constructed on the fly during the simulation to reduce or even wholly flatten free energy barriers. We can trace the idea of biased sampling to the original umbrella sampling method introduced in 1977~\cite{torrie_nonphysical_1977}. The main difference between CV-based enhanced sampling methods lies in how they construct the bias potential and which kind of biased sampling is obtained. Some examples of methods that fall into the category of CV-based enhanced sampling techniques are
local elevation~\cite{huber_local_1994},
adaptive biasing force~\cite{Darve-JCP-2001,Comer2015_TheAdaptiveBiasing,Lesage2016_Smoothed},
energy landscape paving~\cite{Hansmann-PRL-2002},
multiple windows umbrella sampling~\cite{Kastner2011umbreallsampling},
Gaussian-mixture umbrella sampling~\cite{Maragakis-JPCB-2009},
nonequilibrium umbrella sampling~\cite{Warmflash_NoneqUmbrellaSampling_2007,DicksonDinner_NonequilibriumEnhanced_2010},
metadynamics~\cite{laio_escaping_2002,barducci_welltempered_2008,valsson_enhancing_2016},
metabasin metadynamics~\cite{DamaHocky_Metabasin_2015},
parallel-bias metadynamics~\cite{Pfaendtner_PBMetaD_2015},
basis function sampling~\cite{Whitmer_BFS_2014},
Green's function sampling~\cite{Whitmer_GFS_2015},
artificial neural network sampling~\cite{Sidky_ANNSampling_2018},
reweighted autoencoded variational Bayes for enhanced sampling~\cite{Tiwary_RAVE_2018},
on-the-fly probability-enhanced sampling~\cite{Invernizzi2020opus,invernizzi2020unified},
adaptive topography of landscapes for accelerated sampling~\cite{giberti2021atlas},
and reweighted Jarzynski sampling~\cite{Bal_RewieghtedJarzynski_2021}.

Variationally Enhanced Sampling (VES)~\cite{valsson_variational_2014} is a recently developed CV-based enhanced sampling method based on a variational principle. It introduces a convex functional of the bias potential that is related to the relative entropy and the Kullback-Leibler divergence~\cite{valsson_variationally_2018}. To minimize the functional, we generally take the bias potential as a linear expansion in some basis function set. Bias potentials based on neural network~\cite{bonati_neural_2019} or free energy models~\cite{piaggi_variational_2016,McCarty_bespoke_2016,invernizzi_coarse_2017,invernizzi_making_2019} have also been considered in the literature. VES not only allows for obtaining FESs but can also be used to obtain kinetic properties~\cite{mccarty_variationally_2015}.

The focus of this paper is the choice of basis set in the linear expansion of the bias potential within VES\@. So far, the basis functions employed have been primarily global functions such as plane waves, Chebyshev, or Legendre polynomials that are orthogonal but delocalized in the CV space. Gaussian basis functions have also been used~\cite{demuynck_efficient_2017,demuynck_protocol_2018}. However, there has not been an extensive study of how the choice of the basis functions affects the convergence behavior. In particular, it remains an open question if basis functions that are localized in the collective variable space might perform better. While Gaussian basis functions might be the type of localized basis functions that first comes to mind, they have the disadvantage of not forming orthogonal basis sets. Instead, a more appealing option might be Daubechies wavelet-based basis sets~\cite{daubechies_orthonormal_1988}, as they are orthogonal and exhibit an attractive multiresolution property. Daubechies wavelets have recently been used as basis functions for other applications within molecular simulations, such as density functional theory~\cite{mohr_daubechies_2014,Ratcliff_BigDFT_2020} or coarse-grained potentials\cite{maiolo_wavelets_2015}.

In this work, we introduce the Daubechies wavelets as basis functions for the variationally enhanced sampling method.
We implement the wavelets into
into the PLUMED 2 code~\cite{tribello_plumed_2014}, tune their parameters, and evaluate their performance on various systems, going from model potentials to the calcium carbonate association process in water~\cite{kellermeier_entropy_2016}. We also test Gaussians and cubic B-splines as other types of localized basis functions.
Section~\ref{sec:theory} presents the theory of the VES method and introduces the new basis functions.
Besides the theoretical properties, we also provides details on the implementation of the new functionality into the VES module of PLUMED 2~\cite{tribello_plumed_2014}.
In Section~\ref{sec:application}, we present the computational details of the benchmark systems.
We discuss the results of the simulations in Section~\ref{sec:discussion}, and in Section~\ref{sec:summary} we end with some concluding remarks.

\section{Theory and Methodology}\label{sec:theory}

\subsection{CV-based Enhanced Sampling}
We consider a molecular system described by the set of atomic coordinates $\vec{\pmb{r}}$ and a potential energy function $U(\vec{\pmb{r}})$. Without the loss of generality, we limit our discussion to the canonical (NVT) ensemble in the following.
The Boltzmann distribution, which we want to sample by molecular dynamics (MD) or Monte Carlo simulations, is defined as
\begin{equation}
P(\vec{\pmb{r}}) =
\frac{\mathrm{e}^{-\beta U(\vec{\pmb{r}})}}
{\int \mathrm{d} \vec{\pmb{r}}\, \mathrm{e}^{-\beta U(\vec{\pmb{r}})}}
\end{equation}
where $\beta = (k_\mathrm{B} T)^{-1}$ is the inverse of the thermal energy.
In collective variable (CV) based enhanced sampling methods, we identify a few relevant CVs that correspond to critical slow degrees of freedom.
The equilibrium probability distribution corresponding to a set of CVs, $\pmb{s} (\vec{\pmb{r}})= \{s_1(\vec{\pmb{r}}),s_2(\vec{\pmb{r}}),\ldots,s_N(\vec{\pmb{r}})\}$,
is given by
\begin{equation}
\label{eq:equ_CV}
P(\pmb{s}) =
\int \mathrm{d} \vec{\pmb{r}}\, \delta(\pmb{s} - \pmb{s}(\vec{\pmb{r}}))  P(\vec{\pmb{r}}) =
\langle \delta(\pmb{s} - \pmb{s}(\vec{\pmb{r}})) \rangle
\end{equation}
while the free energy surface (FES) is defined as
\begin{equation}
\label{eq:fes_CV}
F(\pmb{s}) = - \beta^{-1} \log P(\pmb{s}) + C
\end{equation}
where $C$ is an additive constant.

We are generally interested in systems where the FES (or equivalently the equilibrium probability distribution $P(\pmb{s})$) is hard to sample by unbiased molecular dynamics simulations. For example, the FES might be characterized by many metastable basins separated by high free energy barriers such that barrier crossings occur on far greater time scales than we can afford in simulations, that is, are rare events.

To overcome this time scale or rare event problem, we can enhance the sampling by introducing a bias potential $V(\pmb{s}(\vec{\pmb{r}}))$ that acts in the space of the CVs. The introduction of this bias potential will lead to a biased (i.e., non-Boltzmann) distribution  given by
\begin{equation}
P_{V}(\vec{\pmb{r}}) =
\frac{\mathrm{e}^{-\beta \left[U(\vec{\pmb{r}})+V(\pmb{s}(\vec{\pmb{r}}))\right]}}
{\int \mathrm{d} \vec{\pmb{r}} \,
\mathrm{e}^{-\beta \left[U(\vec{\pmb{r}})+V(\pmb{s}(\vec{\pmb{r}}))\right]}}
\end{equation}
Consequently, this leads to a biased CV distribution given by
\begin{equation}
  P_{V}(\pmb{s}) = \int \mathrm{d} \vec{\pmb{r}} \, \delta(\pmb{s} -\pmb{s}(\vec{\pmb{r}}))
  \,
P_{V}(\vec{\pmb{r}})
\propto \mathrm{e}^{-\beta\left[F(\pmb{s})+V(\pmb{s})\right]}
\end{equation}
that is chosen such that the sampling is easier and free energy barriers are reduced or even completely flattened.

From the biased simulation, we can obtain an ensemble average of an observable $O(\vec{\pmb{r}})$ for the unbiased simulation through reweighting
\begin{equation}
\label{eq:reweighthing}
\langle O(\vec{\pmb{r}}) \rangle =
\frac{\langle O(\vec{\pmb{r}}) \, w(\vec{\pmb{r}}) \rangle_{V}}
{\langle w(\vec{\pmb{r}}) \rangle_{V}}
\end{equation}
where $w(\vec{\pmb{r}})=e^{\beta V(\pmb{s}(\vec{\pmb{r}}))}$ is the weight of configuration $\vec{\pmb{r}}$ and the averages on the right side are obtained in biased ensemble. In particular, we can obtain the FES for some CV set $\pmb{s}'$ by using $O(\vec{\pmb{r}}) = \delta(\pmb{s}' - \pmb{s}'(\vec{\pmb{r}}))$
\begin{equation}
\label{eq:reweighthed_fes}
F(\pmb{s}) = - \beta^{-1} \log \, \langle \delta(\pmb{s}' - \pmb{s}'(\vec{\pmb{r}})) \, w(\vec{\pmb{r}}) \rangle_{V} + C'
\end{equation}
where we can ignore the denominatior in eq~\ref{eq:reweighthing} as it only gives a constant shift of the FES (i.e., we can include it in the constant $C'$). In practice, the reweighted FES is obtained using a reweighted histogram or kernel density estimation where each sample is weighted by the bias acting on it, $w(\vec{\pmb{r}})=e^{\beta V(\pmb{s}(\vec{\pmb{r}}))}$. The reweighting procedure of eq~\ref{eq:reweighthing} assumes a fixed bias potential, but often it can be used for adaptively constructed bias potential under the assumption that the bias potential is quasi-stationary, as we discuss below.

\subsection{Variationally Enhanced Sampling}
In the VES method introduced by Valsson and Parinello~\cite{valsson_variational_2014}, the bias potential is constructed by minimizing a convex functional given by
\begin{align}
  \Omega[V] =
  \frac{1}{\beta}\log \frac
    {\int \mathrm{d} \pmb{s}\, \mathrm{e}^{-\beta \left[ F(\pmb{s}) + V(\pmb{s}) \right]}}
    {\int \mathrm{d} \pmb{s}\, \mathrm{e}^{-\beta F(\pmb{s})}}
    + \int \mathrm{d} \pmb{s}\, p(\pmb{s}) V(\pmb{s})
\end{align}
where $p(\pmb{s})$ is a normalized probability distribution.
The stationary point of this functional is given up to a constant by
\begin{align}
  V(\pmb{s}) = - F(\pmb{s}) - \frac{1}{\beta} \log p(\pmb{s})
  \label{eq:func_stat}
\end{align}
which, due to the convexity of $\Omega[V]$, is the global minimum.
At this minimum, the CVs are distributed according to $p(\pmb{s})$, which is consequently called a ``target distribution''.
It can be shown that the $\Omega[V]$ functional is related to the Kullback-Leibler divergence (or relative entropy) and the cross entropy~\cite{valsson_variationally_2018}.

Thus, by minimizing $\Omega[V]$, we can construct a bias potential that leads to a sampling of the CVs according to the target distribution  $p(\pmb{s})$. The most straightforward choice of the target distribution is a uniform target distribution, leading to completely flat sampling in CV space. However, we have found it better to employ a so-called well-tempered target distribution~\cite{barducci_welltempered_2008,valsson_welltempered_2015} given by $p(\pmb{s}) = \left[P(\pmb{s}) \right]^{1/\gamma} / \int \mathrm{d} \pmb{s} \, \left[P(\pmb{s}) \right]^{1/\gamma}$, where $\gamma$ is a parameter, named bias factor, that determines how much the sampling is enhanced as compared to the equilibrium distribution $P(\pmb{s})$.

We can determine the FES directly from the bias potential through eq~\ref{eq:func_stat}. Alternatively, we can obtain the FES, both for the biased CVs and also for any other set of CVs, by using the reweighting procedure shown in eq~\ref{eq:reweighthing}. While the VES bias potential is time-dependent, it quickly becomes quasi-stationary. Therefore, this reweighting procedure is valid after a short intial transient in the time series that is ignored. Note that differently from metadynamics~\cite{tiwary_timeindependent_2015,valsson_enhancing_2016}, we generally do not need to account for time-dependent constants when performing reweighting with VES. Furthermore, under certain conditions, the VES method can also be used to obtain kinetic properties~\cite{mccarty_variationally_2015}.

In practice, we perform the minimization of the $\Omega[V]$ functional by assuming a functional form of the bias potential $V(\pmb{s}; \pmb{\alpha})$ that depends on a set of variational parameters $\pmb{\alpha} = \{\alpha_1,\alpha_2,\ldots,\alpha_M\}$. Thus, we go from an abstract functional minimization to a minimization of the multi-dimensional function $\Omega(\pmb{\alpha})$.

Most general is to take the bias potential to be a linear expansion in some set of basis functions $\pmb{f} = \{f_1,f_2,\ldots,f_M\}$,
\begin{align}
  \label{eq:linear_expansion1}
  V(\pmb{s}; \pmb{\alpha}) = \sum_i \alpha_i \, f_i(\pmb{s})
\end{align}
We can then obtain the gradient $\nabla\Omega(\pmb{\alpha})$ and the Hessian  $H_{\Omega}(\pmb{\alpha})$ as
\begin{align}
  \nabla\Omega(\pmb{\alpha})_{i} =
  \frac{\partial \Omega(\pmb{\alpha})}{\partial \alpha_i} =
    - \left\langle f_i(\pmb{s})\right\rangle_{V(\pmb{\alpha})}
    + \left\langle f_i(\pmb{s}) \right\rangle_p  \label{eq:gradient}\\
  H_{\Omega}(\pmb{\alpha})_{i,j} =
  \frac{\partial^2 \Omega(\pmb{\alpha})}{\partial \alpha_i \alpha_j} =
    \beta\, \mathrm{Cov}[ f_j(\pmb{s}), f_i(\pmb{s})]
      _{V(\pmb{\alpha})}
\end{align}
where angular brackets denote expectation values and $\mathrm{Cov}[\ldots]$ the covariance, obtained either over the bias potential or over the target distribution.

Due to statistical sampling, the estimates of the gradient and Hessian are generally noisy. Therefore, we perform the minimization of $\Omega(\pmb{\alpha})$ using stochastic optimization algorithms. In particular, the averaged stochastic gradient descent algorithm from ref~\citenum{bach_nonstronglyconvex_2013} has proven a convenient choice. In this algorithm, the instantaneous parameters are updated according to the following recursive equation
\begin{align}
  \pmb{\alpha}^{(n+1)} = \pmb{\alpha}^{(n)}
  - \mu \left[\nabla\Omega(\bar{\pmb{\alpha}}^{(n)}) + H_{\Omega}(\bar{\pmb{\alpha}}^{(n)})
  (\pmb{\alpha}^{(n)} - \bar{\pmb{\alpha}}^{(n)}) \right]
  \label{eq:bach_sgd}
\end{align}
where $\mu$ is a constant step size and the gradient and Hessian are obtained using the averaged parameters $\bar{\pmb{\alpha}}^{(n)} = \frac{1}{n+1} \sum_{i=0}^{n} \pmb{\alpha}^{(i)}$ (i.e., the bias potential depends on the averaged parameters). The parameters are updated with a relatively small stride, on the order of 1000 MD steps. Here, we only employ the diagonal part of the Hessian matrix, as generally done in VES~\cite{valsson_variational_2014,valsson_variationally_2018}.

\subsection{Linear Basis Functions for VES}

The focus of this paper is the basis functions used in the linear expansion of the bias potential (eq~\ref{eq:linear_expansion1}). So far, the basis functions employed have been global functions such as plane waves (i.e., Fourier series)~\cite{valsson_variational_2014}, Chebyshev polynomials~\cite{valsson_welltempered_2015}, or Legendre polynomials. The usage of global functions is closely related to the idea of using spectral methods for function approximation.~\cite{boyd_chebyshev_2001}. Favorable for their usage within VES, these basis functions form complete and orthogonal basis sets. However, they are delocalized in the CV space. In other words, they are non-zero over their full domain except on isolated points.

Using global or delocalized basis functions means that during the optimization process the bias potential will change even in parts of CV space where the MD simulation is not currently exploring. While this has not proven to be a significant issue, it is clear that delocalized basis functions might not be the optimal choice.

In this work, we consider the performance of using VES with localized basis functions, that is, functions that are non-zero on only some part of the domain of the bias potential. Therefore, they should not suffer from the issue of the bias potential changing in parts of CV space that the simulation is not currently exploring.

Examples of such localized basis functions that come to mind would be Gaussians or splines. In fact, in refs~\citenum{demuynck_efficient_2017,demuynck_protocol_2018}, the authors employed VES with Gaussian basis functions. The results obtained with this VES setup were found to be inferior to some of the results obtained with other enhanced sampling methods used by the authors (such as umbrella sampling~\cite{torrie_nonphysical_1977}), but as no other basis functions were used with VES, it is hard to judge the performance of the Gaussian basis from their results.
However, one disadvantage with using Gaussians or splines as basis functions is that they do not form orthogonal basis sets, which might affect the optimization process.

We have thus been motivated to explore the usage of wavelets as basis functions. In particular, we consider Daubechies wavelets~\cite{daubechies_orthonormal_1988,daubechies_ten_1992} which are localized functions that form orthogonal and complete basis sets. Furthermore, they have an intrinsic multiresolution property that makes it possible to iteratively add more basis functions on smaller scales in a way that preserves orthogonality of the basis.

\begin{figure}
  \includegraphics{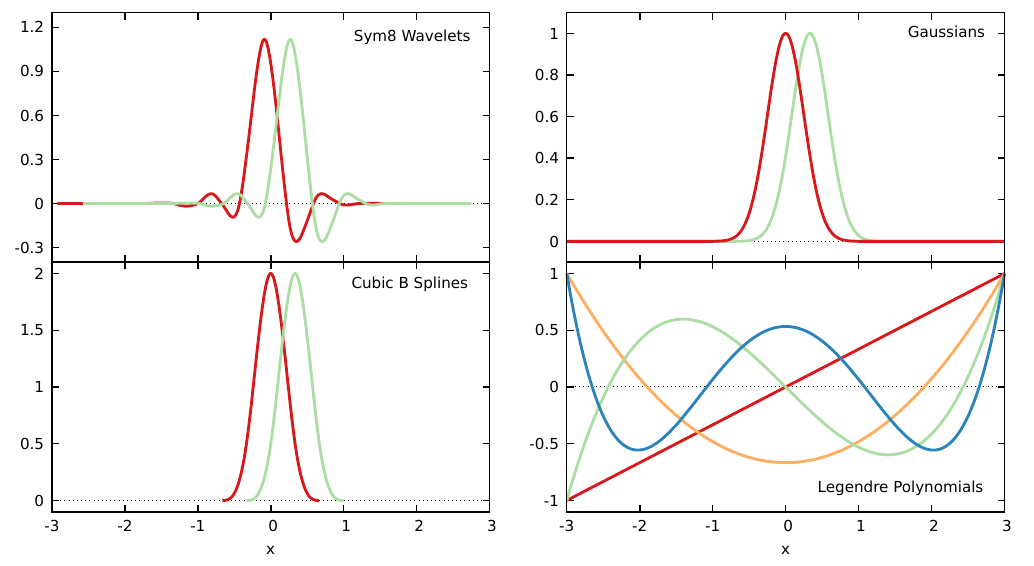}
  \caption{Visualization of different VES basis functions used in this paper.
  The Sym8 wavelets, Gaussians, and cubic B-splines are localized basis functions.
  Here, we only show two adjacent functions while a full basis set would include all shifted functions in the given interval (that is $[-3,3]$ here).
  On the contrary, Legendre polynomials are delocalized functions supported on the full interval of the bias.
  The Legendre basis set consists of all polynomials up to a certain order, the figure shows the functions up to the quartic polynomial.}
  \label{fig:basisfunctions}
\end{figure}

In the following Sections, we briefly describe the new localized basis functions --- Daubechies wavelets, Gaussians and cubic B-splines --- as well as Legendre and Chebyshev polynomials that we consider for comparison. These basis functions are shown in Figure~\ref{fig:basisfunctions}. We give descriptions of one-dimensional basis functions only, as basis sets for higher dimensions can be obtained by considering a tensor product. For example, in two dimensions we obtain
\begin{equation}
  V(s_1, s_2; \pmb{\alpha}) = \sum_{i,j} \alpha_{i,j}\, g_i(s_1)\, h_j(s_2)
\end{equation}
where $g_i(s_1)$ and $h_j(s_2)$ are some one-dimensional basis functions.
All the one-dimensional basis functions described in the following are defined on some given interval $[a,b]$ and include an additional constant basis function. In practice, for MD simulations, we also need the derivatives of the basis functions to obtain the biasing force due to external bias potential, but this is a straightforward task for all of the basis functions considered here.

\subsection{Daubechies Wavelets Basis Functions}

Daubechies developed a theory for special types of wavelets that can be used to construct complete and orthogonal basis functions~\cite{daubechies_orthonormal_1988}. These wavelets are based on using a pair of functions, the scaling function (or father wavelet) $\phi$, and the wavelet function (or mother wavelet) $\psi$.
They are defined by
\begin{align}
  \phi_k^j (x) &= 2^{-j/2} \phi ( 2^{-j} x - k )\\
  \psi_k^j (x) &= 2^{-j/2} \psi ( 2^{-j} x - k )
\end{align}
for a given scale $j \in \mathbb{Z}$ and shift $k \in \mathbb{Z}$.
The exact properties are set by choosing the filter coefficients $h_k$ and $g_k$ in the refinement relations given by:
\begin{align}
  \phi (x) &= \sqrt{2} \sum_k h_k\, \phi ( 2 x - k)\\
  \psi (x) &= \sqrt{2} \sum_k g_k\, \phi ( 2 x - k)
\end{align}

Daubechies proved that certain finite sets of filter coefficients result in orthonormal bases.
Using these wavelet functions, any square-integrable function $g(x)$ can be approximated up to arbitrary precision by a linear combination with coefficients $\alpha$
\begin{align}
  g(x) = \sum_k \alpha_k \, \phi^j_k (x) + \sum_{l \geq j} \sum_{k} \alpha_{l,k} \, \psi^l_k (x)
  \label{eq:wavelet_approx}
\end{align}
where the wavelet functions satisfy orthogonality relations\cite{goedecker_wavelets_1998}:
\begin{align}
  \int \mathrm{d} x \, \phi^j_k(x) \phi^{j}_{k'}(x)   &= \delta_{kk'}\\
  \int \mathrm{d} x \, \phi^j_k(x) \psi^{j'}_{k'}(x)  &= 0 \qquad \qquad \text{for } j \leq j'\\
  \int \mathrm{d} x \, \psi^j_k(x) \psi^{j'}_{k'}(x)  &= \delta_{jj'} \delta_{kk'}
\end{align}
We can see the multiresolution property of the wavelet basis functions in eq~\ref{eq:wavelet_approx}. Starting with the father wavelets $\phi$ at some scale $j$, an increasingly more accurate approximation is obtained by adding mother wavelets $\psi$ at finer scales.

In this paper, we will focus on the coarsest approximation only, which corresponds to a single level of father wavelets at some scale $j$
\begin{align}
  g(x) = \sum_k \alpha_k \, \phi^j_k (x)
  \label{eq:wavelet_approx2}
\end{align}
Left for us to choose are the exact wavelet type and the scale.

The wavelet type is determined by the set of filter coefficients $h_k$ and $g_k$.
Desirable properties for our application are small support of the individual function, at least $C^1$ regularity (one continuous derivative) and the reproduction of polynomials up to a desired order.

The wavelets developed by Daubechies satisfy these properties and in fact result in the minimally supported functions for a given polynomial order.
In this paper we consider filter coefficients that result in the least asymmetric variant of these wavelets or so-called symlets~\cite{daubechies_orthonormal_1988}.
The reduced asymmetry of the symlets comes at the cost of slightly reduced regularity as compared to the conventional maximum phase Daubechies wavelets. However, this does not cause problems as we only require one continuous derivative.
In practice, we found the symlets to perform better than the maximum phase Daubechies wavelets. The symlets are also used in wavelet-based density functional theory calculations~\cite{Ratcliff_BigDFT_2020}. We will denote the symlets by Sym$N$, where $N$ is equal to half the number of coefficients used for construction.

The chosen number $N$ determines the properties of the symlets, including the number of vanishing moments of the mother wavelet. Having $N$ vanishing moments means that all polynomial functions up to order $N-1$ are orthogonal to the mother wavelet. Consequently, any polynomial of order up to $N$ can be be represented exactly by a single level of the father wavelet $\phi$ (i.e., the scaling function). Employing a wavelet basis with a larger $N$ can thus help to construct a bias potential with less regularity and steeper slopes. On the other hand, the range over which the wavelet functions are non-zero is proportional to $2N-1$. Because the basis consists of integer-shifted functions, a larger support (i.e., non-zero range) results in more overlap between functions. This makes it necessary to use more basis functions at the same scale and thus results in more expansion coefficients to optimize. After some testing, we found that using Sym8 or Sym10 yields the best results for the system considered in this paper. Further discussion and a comparison of symlets with different numbers of vanishing moments can be found in Section~S1 of the Supporting Information (SI).

The scale $j$ of the wavelet basis can be chosen freely.
Instead of selecting the scale directly, we set the desired number of basis functions.
In principle, there is an infinite number of shifted wavelet functions in the basis.
However, only a few of them are supported inside the range $[a,b]$ on which the bias potential is defined.
Furthermore they are non-zero only on a small part of their domain.
Thus, we choose to only include the ones with any (absolute) function value inside the bias range that is at least $1\, \%$ of the maximal function value.
We then calculate the required scaling to arrive at the desired number of basis functions.
We did not observe disadvantages from excluding wavelets with minor contributions, while it allows us to reduce the number of coefficients to be optimized.

Generally, using a smaller scale and, consequently, more basis functions allows us to represent finer features better, at the cost of needing to optimize more variational parameters. In Section~S1 of the SI, we show results where we change the number of the basis functions for a fixed $N$ value.

\subsection{Gaussian Basis Functions}
Gaussian basis functions are given by the mathematical expression
\begin{align}
  f_i(x) &= \exp\left(-\frac{{\left(x-\mu_i\right)}^2}{2\sigma^2}\right)
\end{align}
where $\mu_i$ is the center of the individual Gaussian and $\sigma$ is a constant width parameter.
The full basis set is then given by Gaussians functions with centers distributed evenly on the interval $[a,b]$. We add the first center at $\mu_0=a$ and define the shift between centers as $d = \mu_i - \mu_{i-1} = (b-a)/N$, where $N$ is a user-specified integer fixing the number of basis functions.

To mitigate systematic errors at the boundaries, we add one function on each side outside the range, resulting in a total of $N+3$ basis functions including the constant. As the force from the VES bias is zero outside the chosen interval by design, these additional functions will only contribute inside the bias range, similarly to the boundary correction approach for Metadynamics in ref~\citenum{baftizadeh_protein_2012}. Although more complicated boundary correction algorithms have been developed~\cite{crespo_metadynamics_2010,mcgovern_boundary_2013}, we found our simple approach to work well.

The width $\sigma$ of the Gaussians is set by the user. For a fixed number of Gaussians, the possible resolution of the basis can be increased by choosing Gaussians with a smaller width. However, reducing the width will reduce the overlap between Gaussians and a too-small width will result in an ill-behaving basis set. Thus, the optimal width, which very likely is system dependent, is the smallest one that still results in good convergence.
In refs~\citenum{demuynck_efficient_2017,demuynck_protocol_2018} the width $\sigma$ was set equal to the distance $d$ between the centers of the Gaussians. However, as shown in Section~S2 in the SI, we found improved performance when using a smaller width of $\sigma = 0.75d$. Because this yielded better results for the model systems considered here, we will show only Gaussian results obtained with this optimal width in the rest of the paper, while we refer the reader to the SI for results obtained with other $\sigma$ values.

\subsection{Cubic B-Splines Basis Functions}
We consider the cubic B-spline basis functions from ref~\citenum{habermann_multidimensional_2007} that are given by the mathematical expression
\begin{align}
  f_i(x) = h\left(\frac{x-\mu_i}{\sigma}\right)
\end{align}
where
\begin{align}
  h(t) =
  \begin{cases}
    (2 - \lvert t \rvert)^3, & 1 \leq \lvert t \rvert \leq 2\\
    4 - 6\lvert t \rvert^2 + 3 \lvert t \rvert^3,\qquad & \lvert t \rvert \leq 1\\
    0, & \text{elsewhere}
  \end{cases}
\end{align}
and $\mu_i$ is the center of the cubic B-spline basis function and $\sigma$ is the width. The full basis set is then given by spline functions with centers distributed evenly on the interval $[a,b]$. The first center is set on the left boundary $\mu_0 = a$ and we define the shift between centers as $d = \mu_i - \mu_{i-1} = (b-a)/N$, where $N$ is a user-specified integer fixing the number of basis functions. Similar to the Gaussian basis functions, to avoid boundary effects, we add functions on each side outside the range, resulting in a total of $N+3$ basis functions including the constant. Differently from the Gaussians, the width $\sigma$ is fixed and taken as equal to the distance between centers, $\sigma=d$.

\subsection{Legendre and Chebyshev Polynomial Basis Functions}

Legendre and Chebyshev polynomials form sets of orthogonal basis functions on a closed interval that is matched to the range of the bias potential.
Contrary to the previously described bases, the basis functions are not localized in a specific part of the interval but are non-zero except on isolated points.
Chebyshev polynomials of the first kind are given by the recurrence relations
\begin{align}
  C_{0}(x)   &=  1 \\
  C_{1}(x)   &=  x \\
  C_{n+1}(x) &= 2x\,C_{n}(x) - C_{n-1}(x)
\end{align}
while the recursive relations of the Legendre polynomials are
\begin{align}
  L_{0}(x)   &=  1 \\
  L_{1}(x)   &=  x \\
  L_{n+1}(x) &= \frac{2n+1}{n+1} \, x \, L_{n}(x) -  \frac{n}{n+1} \, L_{n-1}(x)
\end{align}
Both Chebyshev and Legendre polynomials are defined intrinsicaly on the interval $[-1,1]$ and need to be scaled and shifted when employed on different intervals.
For a given interval $[a,b]$, we use the following function to transform $t \in [a,b]$ to $x \in [-1,1]$:
\begin{equation}
  x(t) = \frac{2t-(a+b)} {(b-a)}
\end{equation}

\subsection{Implementation of New Basis Functions}\label{sec:implementation}
We have implemented the new basis functions into the VES module of the PLUMED 2 code~\cite{tribello_plumed_2014,theplumedconsortium_promoting_2019}. Our implementation is publicly available in the offical PLUMED 2 GitHub repository and it is released in version 2.8 of PLUMED.

While it was straightforward to implement Gaussians and splines, wavelets pose the problem of not having an analytic mathematical expression.
Instead, in the beginning of the simulation we generate the wavelets values and derivatives on a grid through an iterative scheme. We then use the grid as a lookup table during the simulation. This means that the computational overhead of using the wavelets is minimal.
To generate the wavelet grid, both for the values and the derivatives, we employ a vector cascade algorithm~\cite{strang_wavelets_1997} that relies on finding eigenvectors of a characteristic matrix and subsequent vector-matrix multiplications to iteratively get values on an increasingly finer spaced grid.
We calculate the exact values on a grid of at least 1000 points and use linear interpolation to obtain in-between values.

As localized functions are non-zero only in a small region of the total CV space, we have to modify the optimization scheme slightly. If there is no sampling in the non-zero region of a basis function during one iteration of the bias potential, the elements of gradient and Hessian corresponding to that basis function are set to zero before updating the variational parameters. This is needed because the gradient elements for these basis functions might still be non-zero due to the average over the target distribution (the second term in eq~\ref{eq:gradient}). Setting them to zero prevents erroneous updates of variational parameters if no sampling of the non-zero region occurred. Note that this procedure is done only for individual elements, so the total gradient vector and Hessian matrix still include non-zero elements.

We note that our implementation of the wavelet, Gaussian, and spline basis functions also supports periodic CVs. Furthermore, in addition to the least asymmetric wavelets (i.e., symlets) that we use in this work, the wavelet implementation supports also conventional maximum phase Daubechies wavelets. However, we found the latter to perform worse when compared to the Symlets.

\section{Computational Details}\label{sec:application}
To evaluate the performance of the different basis functions, we perform simulations on different systems, going from model potentials in one and two dimensions, to a realistic system of the association process of calcium with carbonate in water.

\subsection{Double-Well Potential}
\label{sec:compdetail_dw}
We start by considering a single particle moving in a one-dimensional model potential given by
\begin{equation}
  U(x) = x^4 - 4x^2 + 0.7x
\end{equation}
that has two states separated by a barrier of around 5 energy units.
The form of this potential can be seen in Figure~\ref{fig:1dpot_err}\textbf{a}.
We take the $x$-coordinate as the CV such that the reference FES will be given by the potential above, $F(x)=U(x)$ (up to an additive constant).
We employ the \texttt{ves\_md\_linearexpansion} command line tool from the VES code for the simulations. The \texttt{ves\_md\_linearexpansion} tool implements a simple molecular dynamics integrator with a Langevin thermostat~\cite{bussi_accurate_2007}.
We use a time step of 0.005 and a friction coefficient of 10 for the Langevin thermostat.
We set the temperature to $T = 0.5/k_{\mathrm{B}}$, such that the barrier height is about 10 $k_{\mathrm{B}} T$ ($k_{\mathrm{B}}=1$).
We choose to run simulations with four different basis sets: Sym8 wavelets, Gaussians, cubic B-splines, and Legendre polynomials.
We expand the bias potential in the interval from -3 to 3 and fix the number of basis functions to 22 for each basis set to allow for a fair comparison.
We employ a uniform target distribution and update the coefficients of the bias potential every 500 steps.
The stepsize $\mu$ in the averaged stochastic gradient descent optimization algorithm (eq~\ref{eq:bach_sgd}) was adjusted to yield the fastest convergence for each basis set.
We set it to $\mu = 0.5$ for simulations using localized basis functions and decrease it to $\mu = 0.1$ for the simulations with Legendre polynomials.
Each simulation is run for $5 \times 10^6$ steps, while the FES was determined every $5 \times 10^4$ steps via eq~\ref{eq:func_stat}.
For each basis set, we run 20 independent simulations that are started in the global minimum with different random seeds for the initial velocities and random forces.

\subsection{Wolfe-Quapp Potential}
\label{sec:compdetail_wq}
The second model potential is the two-dimensional Wolfe-Quapp potential~\cite{wolfe_chemical_1975,quapp_growing_2005}
\begin{equation}
  U(x,y)= x^4 +y^4 - 2 x^2 -4 y^2 + xy + 0.3 x + 0.1 y
\end{equation}
that has two states separated by a high barrier along the $y$-coordinate, while along the $x$-coordinate the mobility is high.
The potential can be seen in Figure~\ref{fig:2dpot} along with projections on the $x$- and $y$-coordinates.
We take both the $x$-coordinate and the $y$-coordinate as CVs, such that the reference FES will be given by the potential, $F(x,y)=U(x,y)$ (up to an additive constant).
We bias both CVs in the interval from -3 to 3 using 22 basis functions per CV (484 two-dimensional basis functions in total).
We set the temperature to $T = 1 / k_{\mathrm{B}}$. We set the stepsize for all simulations to $\mu = 0.5$.
We run 20 independent simulations for each basis set.
Otherwise, we employ the same basis functions and simulation parameters as for the one-dimensional potential in the previous section.

\subsection{Rotated Wolfe-Quapp Potential}
\label{sec:compdetail_rotated_wq}
To test the behavior when biasing only a suboptimal CV, we consider a rotated and scaled version of the Wolfe-Quapp potential.
As in ref~\citenum{invernizzi_making_2019}, the potential is rotated by an angle of $\theta = -0.15\pi$.
The potential energy surface is given in Figure~\ref{fig:rot_wq} together with projections on the $x$- and $y$-coordinates.
We take only the $x$-coordinate as a biased CV, which results in missing orthogonal slow degrees of freedom (the $y$-coordinate). The reference FES for the $x$-coordinate can be obtained by integrating over the $y$-coordinate, $F(x) = -\beta^{-1} \log \int \mathrm{d} y\, e^{-\beta U(x,y)}$.
We use a temperature of $T = 1 / k_{\mathrm{B}}$.
We expand the bias potential in the interval from -3 to 3 and fix the number of basis functions to 22 for each basis set.
We employ a uniform target distribution and update the coefficients of the bias potential every 500 steps.
Otherwise, we employ the same basis functions and simulation parameters as for the previous two model potentials.

For this system, we observe that using the averaged stochastic gradient descent
optimization algorithm does not yield good convergence for the localized basis functions.
Therefore, we use the Adam stochastic gradient descent algorithm~\cite{kingma_adam_2015}, which has been used previously for VES in combination with neural networks~\cite{bonati_neural_2019}.
Details of the Adam algorithm can be found in Section~S3 in the SI\@.
We notice a high sensitivity of the convergence to the stepsize $\eta$ of the Adam algorithm.
Although the standard value of $\eta = 0.001$ works in most cases, the convergence of the bias is slow, especially for simulations with Sym8 wavelets.
Increasing it to $\eta = 0.005$ provides much better behavior, whereas increasing it even further results in non-converging simulations with Legendre polynomials.
We use $\eta = 0.005$ for all simulations with the Adam algorithm but note explicitly that the choice of parameters seems crucial for good convergence.

While the usage of the Adam algorithm helps improve the convergence for this system, we find worse performance in comparison to the averaged stochastic gradient descent algorithm when testing it on the other systems considered in this paper.
Therefore, further investigation is needed to understand the optimal choice for stochastic optimization.
The choice very likely depends on the form of the bias potential (e.g., a linear expansion versus a neural network~\cite{bonati_neural_2019} or a bespoke model~\cite{piaggi_variational_2016,McCarty_bespoke_2016,invernizzi_coarse_2017,invernizzi_making_2019}) and the basis functions used.
An interesting idea might be to combine ideas from different algorithms, similar as was done in ref~\citenum{invernizzi_making_2019} where the authors introduced a combination between AdaGrad and Bach's algorithms.
However, detailed investigation of the stochastic optimization algorithm used within VES are beyond the scope of the current work.

\subsection{Calcium Carbonate Association}\label{sec:compdetail_caco3}

To study the performance of wavelet basis functions for a realistic system, we consider the association process of a calcium carbonate ion-pair in water.
We use the LAMMPS code\cite{plimpton_fast_1995} (5Jun2019 release) interfaced with the PLUMED 2 code for the simulations. We employ the calcium carbonate force field developed in refs~\citenum{Demichelis_NatComm2011,raiteri_thermodynamically_2015} and the SPC/Fw~\cite{Wu_SPCFw_JCP2006} water model. We follow the computational setup used in a previous metadynamics study of the association process~\cite{kellermeier_entropy_2016} using this force field.
We set up a system that contains a single Ca\textsuperscript{2+}-- CO\textsubscript{3}\textsuperscript{2-} ion-pair and 2448 water molecules in a periodic cubic box. We equilibrate the system in the NPT ensemble at a constant emperature of 300 K and a constant pressure of 1 bar for 500 ps. All subsequent simulations are performed in the NVT ensemble using a constant temperature of 300 K and a cubic box with side lengths 41.69~\AA{}. We run 5~ns of unbiased MD simulations from which we select in total 75 snapshots that we use as initial configurations for the biased simulations.
We employ a time step of 0.001 ps. All simulations are performed at a constant temperature of 300 K using a Nos\'{e}-Hoover thermostat~\cite{nose_unified_1984,hoover_canonical_1985,tuckerman_liouvilleoperator_2006} with a chain length of 5 and a relaxation time of 0.1 ps.
For the NPT equilibration, we employ a Nosé-Hoover barostat with a relaxation time of 1 ps to keep a constant pressure of 1 bar.
Electrostatic interactions are calculated according to the PPPM method~\cite{hockney_computer_1988} with an accuracy of $10^{-5}$.

We use the same CVs as in ref~\citenum{kellermeier_entropy_2016}, namely the distance between the Ca and C atoms and the coordination number of Ca with water (see Section~S5 in the SI for further details).
As in the original work~\cite{kellermeier_entropy_2016}, we use the technique of multiple walkers~\cite{raiteri_efficient_2006} with 25 walkers running in parallel to improve convergence, where each walker starts from a different initial configuration. We employ Sym10 wavelets or Chebyshev polynomials as basis functions.
For the CV corresponding to the distance between the Ca ion and C atom of the carbonate ion, we use 60 basis functions in the range from 2~\AA{} to 12~\AA{}.
For the CV corresponding to the coordination number, we use 30 basis functions in the range 5 to 9.
The total number of two-dimensional basis functions is then 1200.
Due to usage of multiple walkers, we update the coefficients of the bias potential more frequently or every 10 MD steps (the total number of data points for each iteration is then 250).
We use the averaged stochastic gradient descent optimization algorithm with a step size of $\mu=0.001$ for the Sym10 wavelets.
For simulations with Chebyshev polynomials this does not always result in stable simulations and we use a lower stepsize of $\mu=0.0005$ for these.
We employ a well-tempered target distribution~\cite{valsson_welltempered_2015} with a bias factor of 5, where the target distribution is iteratively updated every 100 bias potential updates (1000 MD steps). We run each walkers for 3~ns, resulting in a cumulative simulation time of 75~ns.

For comparison, we also perform a well-tempered metadynamics (WTMetad)~\cite{barducci_welltempered_2008} simulation using the same setup as in ref~\citenum{kellermeier_entropy_2016}. The bias factor is set to 5. For the Gaussians, we use an initial height of 1 $k_{\mathrm{B}}T$, and widths of 0.2~\AA{} and 0.1 for the distance and coordination number, respectively. We deposit Gaussians every 1~ps (1000 MD steps). For the metadynamics simulations, we also run each walker for 3~ns, resulting in a cumulative simulation time of 75~ns.

To focus the sampling in the part of the configuration space of interest for the association process, we add an artificial repulsive wall at a Ca--C distance of 11~\AA{} in all simulations to prevent the ions from moving further apart.
In practice this is implemented by a harmonic bias of the form $\kappa (x - x_0)^2$ where we set the parameters to $\kappa = 12$~eV and $x_0 = 11$~\AA{}.

To obtain the reweighted FESs, we employ a reweighted kernel density estimation as implemented in PLUMED 2. We use Gaussian kernels with bandwidths of 0.05~\AA{} and 0.05 for the Ca--C distance and coordination number CV, respectively. We ignore the first 200 ps of each walker and use samples obtained every 0.1 ps. For the metadynamics simulations, we use the $c(t)$ reweighting scheme described in refs~\citenum{tiwary_timeindependent_2015,valsson_enhancing_2016}. During the metadynamic simulations, we calculate the time-dependent constant $c(t)$ needed for the biasing weights every time a Gaussian is added using a grid of $275 \times 300$ over the domain $[2,13] \times [3,10]$.

To assess the stability of the simulations, we perform 3 independent runs using different initial configurations for each of the 3 biasing setups (VES with wavelets, VES with Chebyshev polynomials, WTMetaD).

\subsection{Performance Measures}
\label{sec:performance_measures}
To evaluate and compare the performance of the basis functions, we consider two different performance measures: the root mean square error with respect to a reference and the free energy difference between some two metastable states.

To measure the quality of the FES $F(\pmb{s})$ obtained directly from the bias through eq~\ref{eq:func_stat}, we calculate the root mean square (RMS) error of the FES with respect to a reference as done in refs~\citenum{valsson_welltempered_2015,branduardi_metadynamics_2012}. Given some reference FES $F_{\mathrm{ref}}(\pmb{s})$, the RMS error is given by
\begin{equation}
  \label{eq:rms_error}
  \epsilon = \sqrt{
  \frac{\int \mathrm{d}\pmb{s} \left[ F(\pmb{s}) - F_{\mathrm{ref}}(\pmb{s}) \right] ^2 \theta(\nu - F_{\mathrm{ref}}(\pmb{s}))}
  {\int\mathrm{d}\pmb{s}\, \theta(\nu - F_{\mathrm{ref}}(\pmb{s}))}
  }
\end{equation}
where we perform the integration over the full CV space, and $\theta$ is a Heaviside step function such that only regions with a free energy lower than a threshold value $\nu$ are considered. Since the FESs are only determined up to a constant, we shift them by their average value in the region of interest, that is, we use
\begin{equation}
  \label{eq:fes_shift}
  \tilde{F}(\pmb{s}) = F(\pmb{s}) - \int_{\Gamma}  \mathrm{d}\pmb{s} \, F(\pmb{s}) + \int_{\Gamma} \mathrm{d}\pmb{s} \, F_{\mathrm{ref}}(\pmb{s})
\end{equation}
to calculate the error metric in eq~\ref{eq:rms_error}, where $\Gamma$ is taken as the region of CV space where $F_{\mathrm{ref}}(\pmb{s}) \leq 4\;k_B T$.
We set the parameter $\nu = 8\; k_B T$. We consider always an ensemble of multiple independent runs that are initiated with different initial conditions because a single simulation might not be representive~\cite{coveney_calculation_2016,grossfield_best_2019}.
We then compare the mean RMS error as well as the associated standard error of the mean.

Another performance measure we can employ is to calculate the free energy difference $\Delta F_{A,B}$ between two different states~\cite{valsson_enhancing_2016}
\begin{equation}
  \Delta F_{A,B} = F_A - F_B =
  -\frac{1}{\beta}\log\frac{P_A}{P_B} =
  - \frac{1}{\beta}
  \log
  \frac{\int_A \mathrm{d} \pmb{s}\, \exp[-\beta F(\pmb{s})]}
  {\int_B \mathrm{d} \pmb{s}\, \exp[-\beta F(\pmb{s})]}
  \label{eq:delta_F}
\end{equation}
where the domains of integration are the regions in CV space associated with the states $A$ and $B$, respectively.

\subsection{Data Availability}
\label{sec:data}
The data supporting the results reported in this paper are openly available at Zenodo~\cite{pampel_benjamin_2022_5851773} (DOI\@: \href{https://doi.org/10.5281/zenodo.5851773}{\texttt{10.5281/zenodo.5851773}}).
All LAMMPS and PLUMED 2 input files and analysis scripts required to reproduce the results reported in this paper are available on PLUMED-NEST (www.plumed-nest.org), the public repository of the PLUMED consortium~\cite{theplumedconsortium_promoting_2019}, as plumID:22.001 at \url{https://www.plumed-nest.org/eggs/22/001}.

\section{Results and Discussion}\label{sec:discussion}

\subsection{Model Potentials}
\label{sec:results_modelpot}
A common way to test the performance of methodological developments of enhanced sampling methods is to consider the dynamics of a single particle on model potentials that emulate prototypical free energy landscapes. We, therefore, start by considering three model potentials, where we compare the performance of the localized basis functions (Sym8 wavelets, Gaussians, and cubic B-splines) to the delocalized Legendre polynomials that have been used as basis functions within VES so far. For these simulations, we always perform 20 independent runs for each set of basis functions and use the performance measures that we have described in the previous  Section~\ref{sec:performance_measures} to compare the FESs obtained from the bias potential via eq~\ref{eq:func_stat}.

\begin{figure}
  \includegraphics{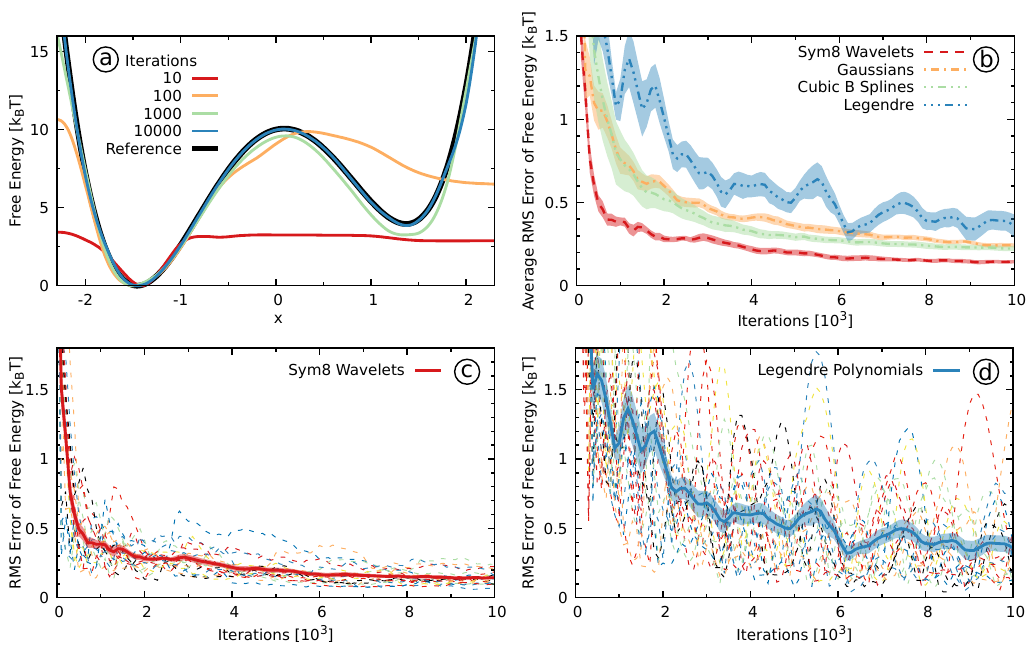}
  \caption{Results for the one-dimensional double-well potential described in Section~\ref{sec:compdetail_dw}.
  (\textbf{a}) The reference FES, along with the FES obtained using the wavelet basis functions at different number of bias iterations for one of the runs.
  (\textbf{b}) The RMS error measure (Section~\ref{sec:performance_measures}, eq~\ref{eq:rms_error}) for the different basis functions as a function of the number of bias iterations. The lines denote the average over 20 independent runs and the shaded area the corresponding standard error.
  (c, d) The RMS error of the individual runs for Sym8 wavelets (\textbf{c}) and Legendre polynomials (\textbf{d}). The thick lines are the same as in (\textbf{b}), the dashed lines each resemble one of the runs.}
  \label{fig:1dpot_err}
\end{figure}

We start by considering the one-dimensional double-well potential shown in Figure~\ref{fig:1dpot_err}\textbf{a} that has a high free energy barrier of around 10 $k_{\mathrm{B}}T$ when going from the left to right side. In panel \textbf{a} of Figure~\ref{fig:1dpot_err}, we show an example of the FES obtained using wavelet basis functions at different bias iterations. In the SI, we present a movie showing the exemplary time evolution of the FES of exemplary simulations for all different basis sets. In panel \textbf{b}, we show the RMS error metric (eq~\ref{eq:rms_error}) for the different basis functions. We can observe that, on average, the FES (or equivalently the bias) converges considerably faster with the localized basis functions than with the delocalized Legendre polynomials. Furthermore, the localized basis functions converge to a better estimate of the FES as indicated by the smaller RMS error. We can observe that the wavelets perform the best of the three localized basis functions.

In Figure~\ref{fig:1dpot_err}\textbf{b}, we can also observe considerably larger fluctuations in the average RMS error and larger standard error for the Legendre polynomials. The reason for this is twofold, as we can see from looking at the RMS error for the individual runs, shown in panels \textbf{c} and \textbf{d} for the wavelets and the Legendre polynomials, respectively. First, within each individual simulation, the bias potential is fluctuating more for the Legendre polynomials. Second, there is a more significant difference between runs for the Legendre polynomials. In comparison, the wavelets show much more robust behavior with considerably smaller fluctuations within individual runs and more minor differences between runs. We can see a similar effect for the Gaussians and cubic B-splines though they do not behave as well as the wavelets (See Figure~S5 in the SI). Therefore, for this simple system, we can already see the benefits of using localized basis functions.

In the following, we will focus on the wavelets and the Legendre polynomials while we refer the reader to the SI for results for the Gaussians and cubic B-splines. Furthermore, we will only use the free energy difference to compare the basis functions while presenting the results for the RMS error metric in the SI\@.

\begin{figure}
  \includegraphics{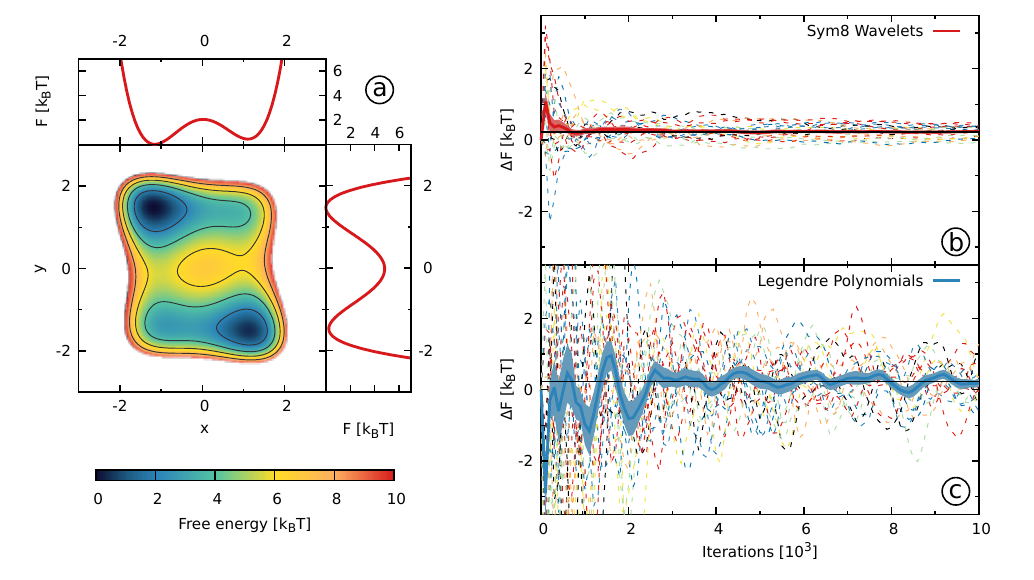}
  \caption{Results for the two-dimensional Wolfe-Quapp potential described in Section~\ref{sec:compdetail_wq}. (\textbf{a}) The reference FES along with free energy projections on the $x$- and $y$-coordinates. (\textbf{b},\textbf{c}) The free energy difference $\Delta F$ (Section~\ref{sec:performance_measures}, eq~\ref{eq:delta_F}) between the two states obtained using Sym8 wavelets (\textbf{b}) and Legendre polynomials (\textbf{c}) as a function of the number of bias iterations. We show results from 20 independent simulations with dashed lines. We use solid lines for the averages and shaded areas to denote the standard errors. We denote the reference value with solid black lines. To define the areas corresponding to the two different states, we use the $y=0$ line.}
  \label{fig:2dpot}
\end{figure}

The next system that we consider is the two-dimensional Wolfe-Quapp potential~\cite{wolfe_chemical_1975,quapp_growing_2005} that is a commonly used model potential for testing methods~\cite{quapp_growing_2005,aguilar-mogas_implementation_2010,bofill_locating_2013,zhang_doubleended_2013}. We show its free energy surface, along with the free energy projections on the $x$- and $y$-coordinates, in Figure~\ref{fig:2dpot}\textbf{a}. The potential has two states separated by a barrier along the $y$-coordinate, while the system is relatively mobile along the $x$-coordinate. Still, due to a strong coupling between the $x$- and $y$-coordinate, it is essential to consider both coordinates as biased CVs to get a good sampling. We thus expand the two-dimensional bias potential in a tensor product basis set of one-dimensional basis functions.

In panels \textbf{b} and \textbf{c} of Figure~\ref{fig:2dpot}, we show the free energy difference between the two states for the wavelets and the Legendre polynomials, respectively. We can see a rather similar behavior as for the one-dimensional model potential. The wavelets exhibit far smaller fluctuations within individual runs and considerably smaller differences between runs than the Legendre polynomial. Looking at averaged free energy difference, we can see that the wavelet simulations converge substantially better and faster than the Legendre polynomials. We can draw similar conclusions by considering the RMS error measure shown in Figure~S7 in the SI.

We show the estimates of the free energy difference from the simulations with Gaussians and cubic B-spline basis functions in Figure~S7 in the SI\@. We can observe that the Gaussians perform better than the Legendre polynomials but worse than the wavelets. However, we find that cubic B-splines perform the worst of all the basis functions and do not yield usable results for this system.

\begin{figure}
  \includegraphics{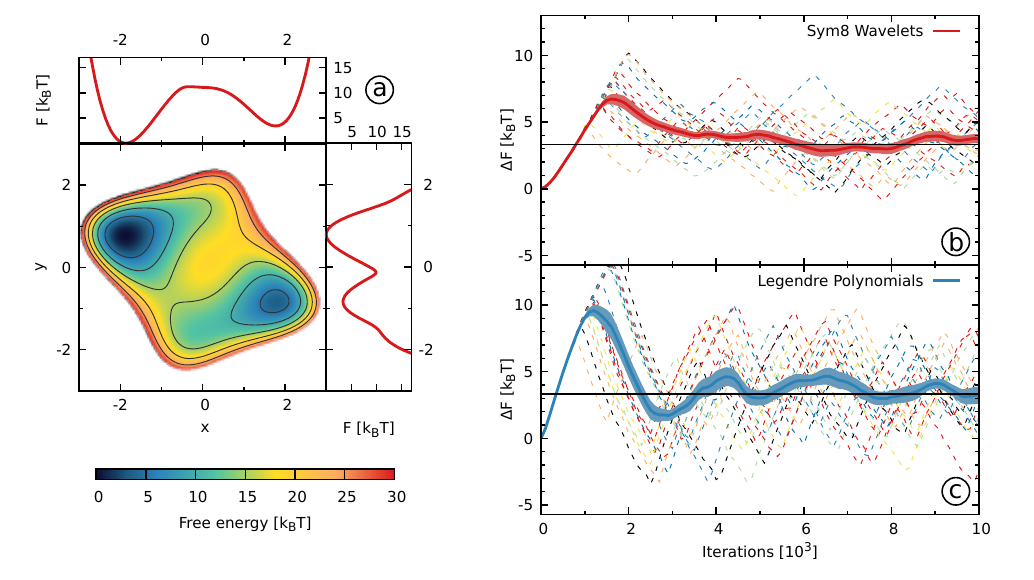}
  \caption{Results for the rotated two-dimensional Wolfe-Quapp potential described in Section~\ref{sec:compdetail_rotated_wq}. (\textbf{a}) The reference FES along with free energy projections on the $x$- and $y$-coordinates. Only the $x$-coordinate is biased. (\textbf{b},\textbf{c}) The free energy difference $\Delta F$ (Section~\ref{sec:performance_measures}, eq~\ref{eq:delta_F}) between the two states obtained using Sym8 wavelets (\textbf{b}) and Legendre polynomials (\textbf{c}) as a function of the number of bias iterations. We show results from 20 independent simulations with dashed lines. We use solid lines for the averages and shaded areas to denote the standard errors. We denote the reference value with solid black lines. To define the areas corresponding to the two different states, we use the $x=0$ line.}
  \label{fig:rot_wq}
\end{figure}

Finally, we consider a rotated version Wolfe-Quapp potential shown in Figure~\ref{fig:rot_wq}\textbf{a} that has been used as a test case for biasing suboptimal CVs~\cite{invernizzi_making_2019,bonati_neural_2019,debnath_gaussian_2020}. We only take the $x$-coordinate as a CV for biasing, so we are missing the $y$-coordinate that is an orthogonal slow degree of freedom. We show the free energy difference between the two states in panels \textbf{b} and \textbf{c} of Figure~\ref{fig:rot_wq}. As expected, due to the usage of a suboptimal CV, the convergence behavior is slightly worse than for the previous two systems, and we need longer simulation times to obtain adequate convergence. Nevertheless, the wavelets exhibit good convergence behavior that, as before, is more robust than for the Legendre polynomials. As shown in Figure~S8 in the SI, the Gaussians and the cubic B-splines perform worse than both wavelets and Legendre polynomials.

As discussed in Section~\ref{sec:compdetail_rotated_wq}, for this system we have used a different optimization algorithm, the Adam optimizater~\cite{kingma_adam_2015} instead of the averaged stochastic gradient descent~\cite{bach_nonstronglyconvex_2013}. This choice might explain a slightly different behavior in the time evolution of individual runs as compared to the previous two systems.

Having tested the localized basis functions on three different model systems, we can draw certain conclusions. The wavelet basis functions exhibit much more robust convergence behavior than the Legendre polynomials. For the wavelets, the fluctuations of the bias potential within individual runs are smaller. Additionally, the difference between independent runs is considerably smaller. The Gaussian and the cubic B-spline basis functions perform worse than the wavelets for all considered systems and do not yield usable results for some systems. Therefore, we recommend against their usage. Having established the excellent performance of the wavelets in model systems, we now move on to their use in a more realistic system.

\subsection{Calcium Carbonate Association}

For a more realistic system, we consider the association process of calcium carbonate in solvent that has previously been investigated in ref~\citenum{kellermeier_entropy_2016} using metadynamics simulations. In that work, the authors used the technique of multiple walkers~\cite{raiteri_efficient_2006} with 25 walkers to improve the convergence. Here, we will follow the same procedure for the wavelet and Chebyshev polynomials simulations. For comparison, we also run well-tempered metadynamics simulations using the same computational setup as used in ref~\citenum{kellermeier_entropy_2016}. For each of the three biasing setups (VES with Sym10 wavelets, VES with Chebyshev polynomials, WTMetaD), we run three independent simulations.

\begin{figure}
  \includegraphics{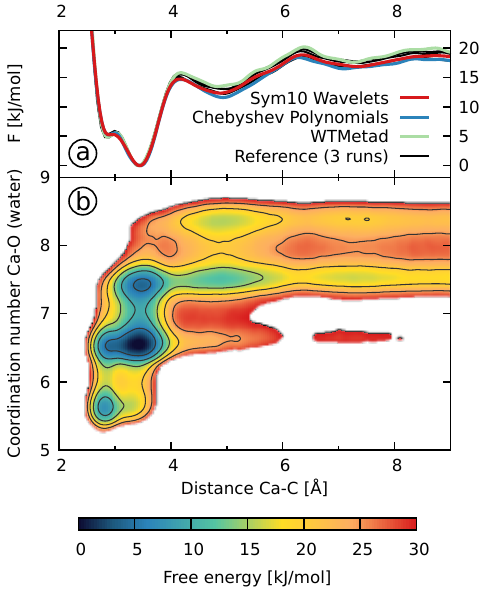}
  \caption{Free energy surfaces for the Calcium Carbonate system described in Section~\ref{sec:compdetail_caco3}.
           (\textbf{a}) Projections on the distance CV for the FESs obtained directly from the bias via eq~\ref{eq:func_stat} (VES) or by summing over the deposited Gaussians (WTMetad).
          We only show one of the runs for each biasing setup.
          The reference data are obtained from ref~\citenum{kellermeier_entropy_2016}.
          (\textbf{b}) FES as a function of both biased CVs obtained by reweighting one of the wavelets simulations.
  }
  \label{fig:CaCO3_FES}
\end{figure}

In Figure~\ref{fig:CaCO3_FES}\textbf{b}, we show the free energy surface as a function of the two biased CVs, the distance between the calcium and the carbon atom of the carbonate and the coordination number of the calcium to the oxygens of the water molecules.
We can see that to fully understand the association process, it is necessary to consider both CVs as the solvation state of the calcium, as measured by the coordination number CV, is closely coupled to the calcium-carbon distance.
The minima of the FES with a Ca--C distance smaller than 4~\AA{} correspond to the states with contact ion-pair.
The lowest state of the FES is the monodentate associated state at around 3.5~\AA{}.
At lower coordination number and smaller distance, a second minimum corresponding to the bidentate state can be seen.
For larger Ca--C distance the ions are no longer in direct contact but are separated by solvent.
The states with a distance of around 5 \AA{} correspond to the solvent-shared ion-pair, while the states around 7 \AA{} denote where the solvation shells of the two ions barely touch.
For even larger distances, the two ions are fully solvated.

To compare the different simulations, we look at the projections of the FES on the distance CV that is shown in Figure~\ref{fig:CaCO3_FES}\textbf{a}. These free energy profiles are obtained at the end of simulations directly from the bias potential, that is via eq~\ref{eq:func_stat} for the VES simulations or by summing up the deposited Gaussians for the WTMetad simulations. For each of the three biasing setups, we only show one representive free energy profile while the profiles for the other runs are shown in Figure~S9 in the SI\@. We also show three reference profiles from ref~\citenum{kellermeier_entropy_2016}. All the free energy profiles are aligned such that their minimum is at zero.

We can observe in Figure~\ref{fig:CaCO3_FES}\textbf{a} that all the free energy profiles obtained from our simulations are in a decent agreement with each other and the reference results from ref~\citenum{kellermeier_entropy_2016}. All the simulations capture reasonably well the small barrier between the mono- and bidentate states at about 3~\AA{}, though we should mention this barrier in the one-dimensional profile does not represent the true barrier of the physical process due to integration over the solvent degree of freedom (i.e., the coordination number CV in the FES shown in panel \textbf{b}). For the dissociated state above 4 \AA{}, we can observe that there are some differences between runs. However, we can observe similar variance between the three reference runs from ref~\citenum{kellermeier_entropy_2016} as shown in Figure~S9 in the SI\@. Therefore, it is difficult for us to say what the correct free energy profile is. Furthermore, our results in panel \textbf{b} are obtained at the end of the simulations and do not reflect that the bias, and thus the obtained FES, fluctuates during the simulation. Indeed, one of the main conclusions from the previous Section~\ref{sec:results_modelpot} was that the fluctuations of the bias potential within individual runs where considerably smaller for the wavelets as compared to the polynomial basis functions.

\begin{figure}
  \includegraphics[width=\columnwidth]{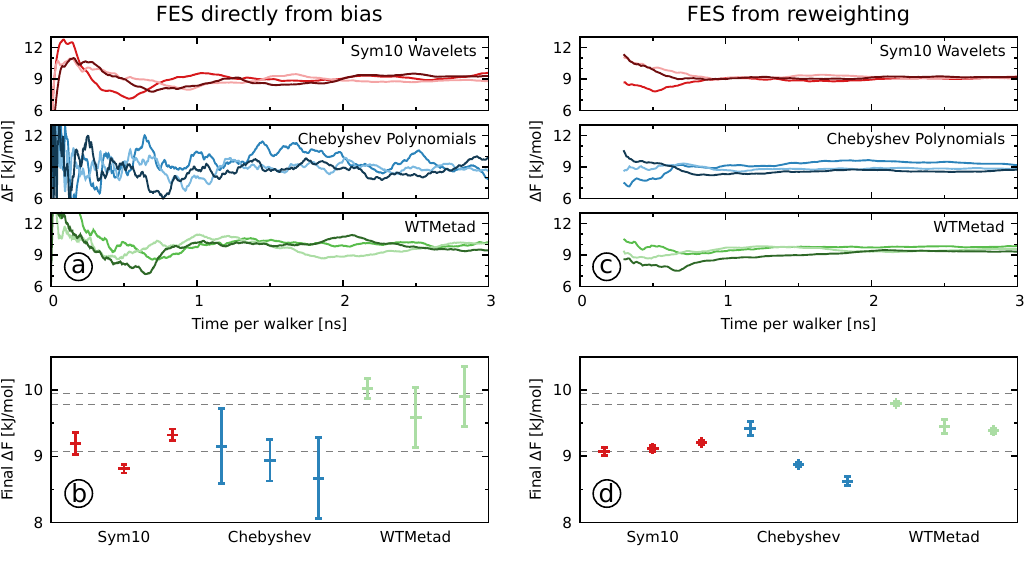}
  \caption{
          Results for the Calcium Carbonate system described in
          Section~\ref{sec:compdetail_caco3}.
          (\textbf{a},\textbf{c}) Time evolution of the free energy difference between the region with Ca-C distance smaller 4~\AA{} and the region with Ca-C distance between 4~\AA{} and 8~\AA{}. For each biasing setup, we show three independent runs where the different color shades represent the individual runs.
          (\textbf{b},\textbf{d}) The average of the free energy differences obtained over the last nanosecond by using 100 samples taken every 10 ps for each simulation. The error bars show the standard deviation to signify the quality of the individual measurements. We also show the results from ref~\citenum{kellermeier_entropy_2016} as black dotted lines. In panels \textbf{a} and \textbf{b}, we use the FES obtained directly from the bias via eq~\ref{eq:func_stat} (VES) or by summing over the deposited Gaussians (WTMetad). In panels \textbf{c} and \textbf{d}, we use the FES obtained through reweighting where we ignore the first 200 ps of each simulation.
  }
  \label{fig:CaCO3_delta_F}
\end{figure}

To gauge the time evolution of the bias potential and FES, we consider the free energy difference between the contact ion-pair and loosely associated states of calcium carbonate. We select the region in CV space with a distance smaller than 4~\AA{} as the contact ion-pair state and the region with distances between 4~\AA{} and 8~\AA{} as loosely associated state and calculate the free energy difference according to eq~\ref{eq:delta_F}. We note that this selection of the two regions does not necessarily coincide with the chemical definitions of ion association states~\cite{kellermeier_entropy_2016}. Here, we employ the free energy difference to monitor the stability of the bias potential and the obtained FES\@. In Figure~\ref{fig:CaCO3_delta_F}\textbf{a}, we show the free energy difference obtained every 10 ps (simulation time per walker). For each of the biasing setups, we show the results from three independent runs.

In Figure~\ref{fig:CaCO3_delta_F}\textbf{a}, we can see that the free energy differences obtained from the wavelet simulations converge faster and show less fluctuations than in the Chebyshev polynomial simulations. In particular, there are considerably larger fluctuations in the Chebyshev polynomial simulations. Furthermore, there is less difference between independent runs for the wavelets as compared to the Chebyshev polynomials. Therefore, when comparing the wavelets and the Chebyshev polynomials, we obtain the same conclusions as for the model system in the previous Section~\ref{sec:results_modelpot}: the wavelets exhibit less fluctuations of the bias potential within individual runs and less difference between different independent runs.
The metadynamics simulations show a convergence behavior that is slightly worse than the wavelet simulations, but still better than the Chebyshev polynomial simulations.

To further quantify the behavior of the simulations, we calculate the average and the standard deviation over the last nanosecond of the each simulations and show the results in Figure~\ref{fig:CaCO3_delta_F}\textbf{b} (numerical values are given in Table S1 in the SI).
We chose the standard deviation because the time series from a single simulation is highly correlated and does not correspond to independent measurements.
The standard deviation is thus shown as a measure of how much the free energy difference and thus also the bias fluctuate even at the end of the simulation.
We can see that there is some spread in the averaged values, though all simulations agree with each other within 1 kJ/mol. We note that there is similar spread in the three reference metadynamics simulations from ref~\citenum{kellermeier_entropy_2016} that are shown as black dotted lines in Figure~\ref{fig:CaCO3_delta_F}\textbf{b}. Therefore, we cannot determine a reference value of the free energy difference. Noticeably, and consistent with the free energy differences time evolution in panel \textbf{a}, the wavelet simulations have the smallest standard deviation values while the values are three to six times larger for the Chebyshev polynomial and metadynamics simulations.

From the results in panels \textbf{a} and \textbf{b} of Figure~\ref{fig:CaCO3_delta_F}, we can conclude that the wavelet perform the best when considering the difference between independent simulations and fluctuations within runs.

So far we have estimated the FES directly from the bias potential. An alternative way to obtain the FES is through reweighting. In fact, it is always a good practice to estimate the FES both directly from the bias potential and via reweighting and compare the results. The reweighting procedure assumes that the bias potential (i.e., the weights) is quasi-stationary. Therefore, we can expect the wavelets to perform better in this respect.

In panel \textbf{c} of Figure~\ref{fig:CaCO3_delta_F}, we show the free energy difference values obtained from reweighted FESs every 10 ps. As before we calculate the average and the standard deviation over the last nanosecond and present it in panel \textbf{d} of Figure~\ref{fig:CaCO3_delta_F}, while numerical values are given in Table S1 in the SI\@.
We can see that there are much smaller fluctuations in the free energy difference for all of the simulations as compared to panel \textbf{b}. All of the wavelet results agree well with each other and when combined yield a numerical estimate of 9.13 $\pm$ 0.04 kJ/mol (see Table S1 in the SI). There is more spread for the Chebyshev polynomial and the metadynamics simulations, but as before all simulations agree within 1 kJ/mol. The reweighted metadynamics values tend to be lower than values obtained directly from the bias potential in panel \textbf{b} and closer to the wavelet results. As for the results obtained directly from the bias potential, we can conclude for the reweighted results that the wavelet perform the best when considering the difference between independent simulations and fluctuations within runs.

Overall for the calcium carbonate association, we find that the wavelet basis functions exhibit excellent performance. The wavelets result in considerably better convergence behavior than the Chebyshev polynomials. The wavelet simulations also show better convergence behavior than the metadynamics simulations.

\section{Conclusions}\label{sec:summary}
In this work, we have introduced the usage of Daubechies wavelets as basis functions for variationally enhanced sampling. We implemented the wavelets into the VES module of the PLUMED 2 code~\cite{tribello_plumed_2014}, have tuned their parameters, and evaluated their performance on model systems and the calcium carbonate association process. Overall, the localized wavelet basis functions exhibit excellent performance and much more robust convergence behavior than the delocalized Chebyshev and Legendre polynomials used as basis functions within VES so far. In particular, the wavelet bases exhibit far smaller fluctuations of the bias potential within individual runs and smaller differences between independent runs. Less fluctuation of the bias potential is important when obtaining FESs and other equilibrium properties through reweighting as the reweighting procedure assumes a quasi-stationary bias potential. Based on our overall results, we can recommend wavelets as basis functions for variationally enhanced sampling.

We have also tested Gaussians and cubic B-splines as other types of localized basis functions. However, the Gaussian and the cubic B-spline basis functions perform worse than the wavelets for all the model systems in Section~\ref{sec:results_modelpot} and do not yield usable results for some systems. Therefore, we recommend against the usage of Gaussians and cubic B-splines as basis functions for VES.

One attractive feature of the wavelets basis functions is the multiresolution property displayed in eq~\ref{eq:wavelet_approx}. Starting with the father wavelets at some given scale, we can  obtain a more accurate approximation of the FES by adding mother wavelets at finer scales. Here, we only employ a single level of father wavelets to expand the bias potential. An interesting future work would be to go beyond this and implement a multiresolution bias potential where we can increase the resolution on the fly during the simulation.
Coupling this with a method to evaluate the quality of the current bias potential on the fly (for example, by using the effective sample size~\cite{ZhangZuckerman_Automated_2010,martino_effective_2017,grossfield_best_2019,Invernizzi2020opus}) could allow us to automatically construct the VES bias potential with a predefined accuracy, without the need to adapt the parameters manually.

Also, in the present work, we focused on a single type of wavelets, the family of Daubechies wavelets in their least asymmetric form. We also initially tested the Daubechies wavelets with extremal phase. However, due to their noticeably worse performance than the Symlets, we did not include them in this work's extensive study.
Nevertheless, other wavelet families could yield better performance for specific systems. Worthwhile to consider might be, for example, the boundary wavelets~\cite{bertoluzza_building_2003} or the multiwavelets developed by Donovan, Geronimo, and Hardin~\cite{donovan_orthogonal_1999,donovan_intertwining_1996}.

\begin{suppinfo}
  \begin{itemize}
    \item PDF file giving further details and results, including: (S1) The effect of the type and scaling parameters for the Daubechies wavelet basis functions. (S2) The effect of the width parameter for the Gaussian basis functions. (S3) The Adam stochastic gradient descent algorithm. (S4) Additional figures for the model potentials. (S5) The collective variables for the calcium carbonate system. (S6) Additional figures for the calcium carbonate system. (S7) Numerical results for the calcium carbonate system.
    \item Video illustrating the time evolution of the FES estimates of the VES method for different basis sets.
  \end{itemize}
\end{suppinfo}

\begin{acknowledgement}
  We thank Paolo Raiteri (Curtin University) for providing the force field for the calcium carbonate system from refs~\citenum{Demichelis_NatComm2011,raiteri_thermodynamically_2015} and providing the reference data from ref~\citenum{kellermeier_entropy_2016}.
  We also thank Stephan Goedecker (University of Basel) for valuable discussions and Robinson Cortes-Huerto and Martin Girard (Max Planck Institute for Polymer Research) for carefully reading over the manuscript.
  We acknowledge support from the Deutsche Forschungsgemeinschaft (DFG, German Research Foundation) - Project number 233630050 - TRR 146 ``Multiscale Simulation Methods for Soft Matter Systems''.
\end{acknowledgement}

\begin{tocentry}
\centering
\includegraphics[width=\columnwidth]{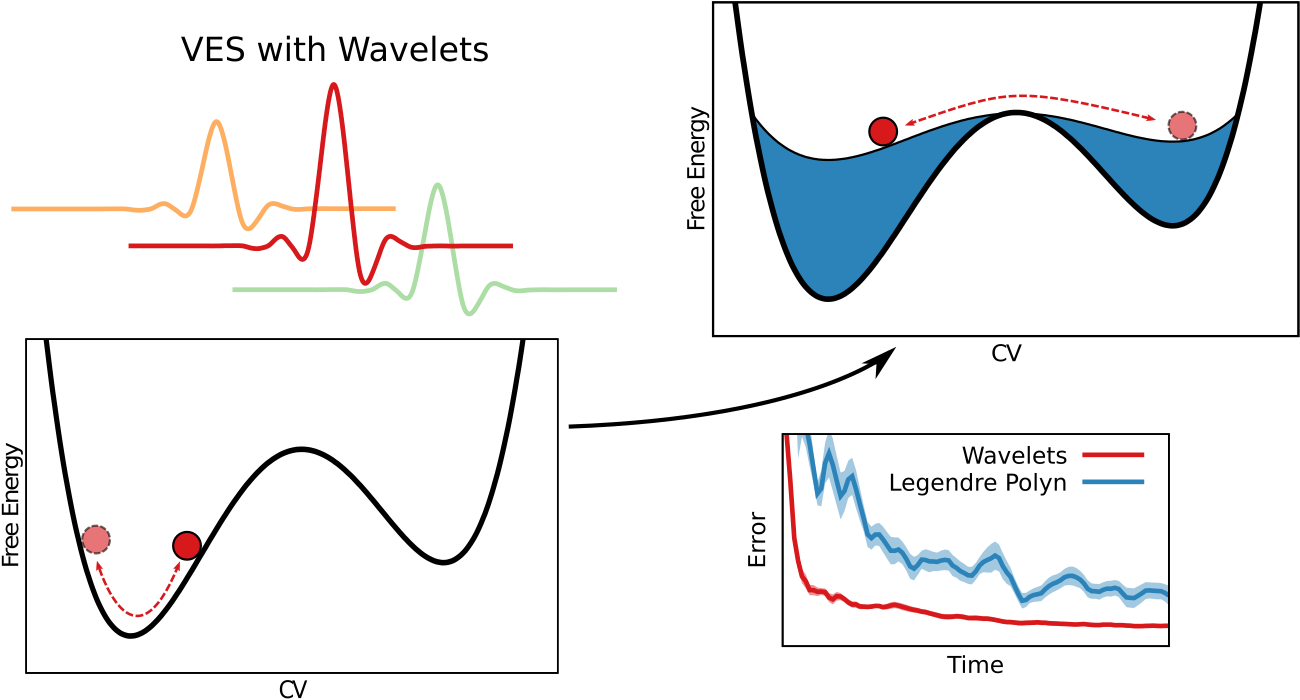}
\end{tocentry}

\bibliography{./WaveletsVES_PampelValsson.bib}

\end{document}